# A Digital Human Model for Symptom Progression of Vestibular Motion Sickness based on Subjective Vertical Conflict Theory

**Shota Inoue, Hailong Liu, and Takahiro Wada**

Human Robotics Laboratory, Nara Institute of Science and Technology, Nara, 630-0192, Japan

**ABSTRACT**

Digital human models of motion sickness have been actively developed, among which models based on subjective vertical conflict (SVC) theory are the most actively studied. These models facilitate the prediction of motion sickness in various scenarios such as riding in a car. Most SVC theory models predict the motion sickness incidence (MSI), which is defined as the percentage of people who would vomit with the given specific motion stimulus. However, no model has been developed to describe milder forms of discomfort or specific symptoms of motion sickness, even though predicting milder symptoms is important for applications in automobiles and daily use vehicles. Therefore, the purpose of this study was to build a computational model of symptom progression of vestibular motion sickness based on SVC theory. We focused on a model of vestibular motion sickness with six degrees-of-freedom (6DoF) head motions. The model was developed by updating the output part of the state-of-the-art SVC model, termed the 6DoF-SVC (IN1) model, from MSI to the MIsery SCale (MISC), which is a subjective rating scale for symptom progression. We conducted an experiment to measure the progression of motion sickness during a straight fore-aft motion. It was demonstrated that our proposed method, with the parameters of the output parts optimized by the experimental results, fits well with the observed MISC.

## INTRODUCTION

The introduction of autonomous driving has amplified the importance of addressing motion sickness owing to the increased potential for motion sickness onset resulting from the increased chances of engaging in a variety of subtasks, including changes in the driving algorithm from human drivers (Diels & Bos, 2016; Sivak & Schoettle, 2015; Wada, 2016). Modeling studies of motion sickness are being actively conducted because the prediction or quantification of sickness is essential to reduce symptoms, while the factors triggering sickness remain unclear.

In the ISO 2631-1, motion sickness dose value (MSDV)(ISO2631-1, 1997), which is defined such that "higher values correspond to a greater incidence of motion sickness" is described as a function of integrating frequency-weighted acceleration in the vertical direction. The MSDV is widely accepted and utilized in a variety of scenarios. However, while MSDV is known to be correlated with the motion sickness incidence (MSI), which is defined as the percentage of people who vomit, its ability to describe milder forms of discomfort or specific symptoms of motion sickness remains unclear. In addition, several mathematical models based on the hypothesis of the mechanism of motion sickness have been developed. Oman (Oman, 1990) was the first to model sensory conflict theory(Reason, 1978), which postulates that the accumulation of discrepancy between afferent neural signals obtained by sensory organs and those expected based on our experiences leads to motion sickness. This model is based on a hypothetical mechanism of human motion perception, which is performed by an internal model thought to be built into the central nervous system (CNS). However, it should be noted that the Oman model is conceptual; as such, the unit or physical meaning of the model output and the source of the conflict are not clearly defined.

The Subjective Vertical Conflict (SVC) theory (Bles et al., 1998) is the only hypothesis in which the source of conflict in SC theory has been clearly defined. SVC theory postulates that the error between the direction of gravity sensed by the sensory organs and the expected direction obtained by the CNS leads to motion sickness. Bos and Bles developed the first mathematical model of the SVC theory for 1 dof vertical motion (Bos & Bles, 1998) based on human motion perception with the idea of the internal model hypothesis. The output of the model is MSI, which was validated by comparing the calculated results with the MSI observed experimentally in (McCauley et al., 1976). Computational models of the SVC theory for 6 dof motion including rotation, have been developed by expanding the internal model framework of motion perception with only OTO in a 1 dof model (Bos & Bles, 1998) to a multisensory structure with the otolith-canal interaction(Inoue et al., 2023; Kamiji et al., 2007; Khalid et al., 2011; Wada et al., 2018). The multisensory integration structure allows us to build several versions of SVC theory models: the inclusion of visual flow(Braccesi et al., 2011)(Wada et al., 2020)(Sousa Schulman et al., 2023; Tamura et al., 2023), static visual vertical(Liu et al., 2024), and the effect of human motion predictability by incorporating hypothetical mechanisms of learning exogenous motion dynamics(Wada, 2021).

As described above, several motion sickness models have been developed; however, MSI is the only model output that can be experimentally measured. Practically, it is desirable to predict or quantify the degree of milder symptoms or the progression of symptoms rather than vomiting. In addition, the onset of motion sickness varies significantly among individuals, although considering such individual differences is important when deriving methods to reduce motion sickness. No model, however, can deal with both of them. Looking beyond the SVC model, there is a model, which outputs the MIsery SCale (MISC)(Bos et al., 2005), an 11-point subjective rating scale based on symptom progression (Irmak et al., 2021). This model (Irmak et al., 2021) uses a 1 DOF lateral acceleration signal instead of the sensory conflict in Oman's SC theory model(Oman, 1990), meaning that the model is not based on the SC theory nor the internal model hypothesis related to motion perception. As such, the model cannot deal with factors that can be addressed in the SVC model families developed thus far, such as multi-DOF motion, including head rotation and motion prediction.

In this context, the purpose of the present study was to develop a motion sickness model to predict symptom progression of motion sickness for six-DoF motions of the head for each individual based on the SVC theory. We developed a model based on (Inoue et al., 2023) among the SVC model families, because it was optimized to fit the observed MSI results and confirmed that it can describe the perceived gravity direction. There are several options for nonlinear mapping from conflict signals related to the discrepancy between the sensed and expected verticalities of MISC. In the present study, we introduce and compare four types of nonlinear dynamics: the one used in the Original SVC model for MSI output(Bos & Bles, 1998; Inoue et al., 2023), the one introduced by Oman for generic sickness severity output(Oman, 1990), and their various combinations. We conducted an experiment to measure the progression of motion sickness during straight fore-aft motion. Based on the experimental results, the parameters of the output parts of the proposed models were optimized to fit the MISC observed for each participant.

## COMPUTATIONAL MODEL OF SYMPTOM PROGRESSION OF MOTION SICKNESS

### MODEL STRUCTURE

Fig.1 shows the structure of the proposed model. The SVC part, which calculates the conflict from the sensory signals from head motions, has the same structure as the SVC models that output the MSI(Inoue et al., 2023). For the output part, which is a dynamic mapping that connects the conflict to the MISC, we propose four different types and compare their performances.

A. SVC part

In (Inoue et al., 2023), the model structure for the error feedback part of the 6DoF-SVC model (Kamiji et al., 2007)(Wada et al., 2018) and the parameters were revisited to investigate the accuracy of self-motion perception for verticality, as well as to increase the accuracy of motion sickness. This study found that the best model was the one in which the integral exists only in the feedback part for the error in the subjective vertical direction, which was referred to as the In-1 model in prior work (Inoue et al., 2023). Therefore, the In-1 model was adopted for SVC.

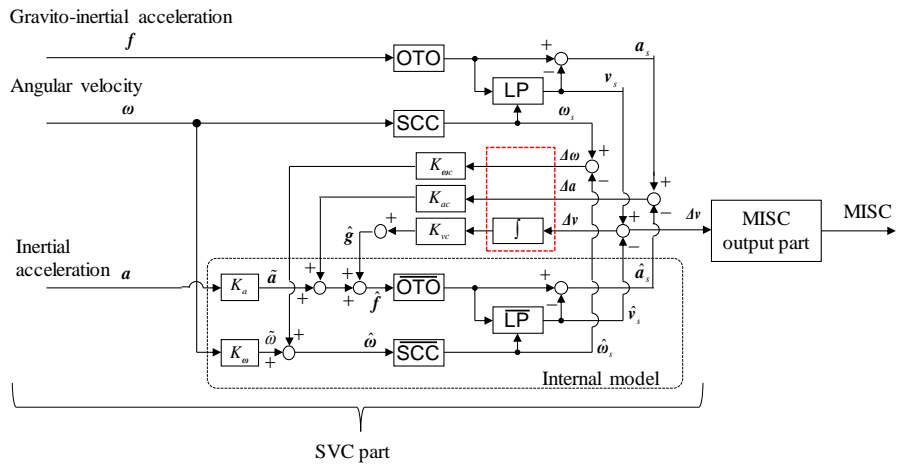

**Figure 1:** Structure of the proposed model. Left hand side of the mode, where subjective vertical conflict Dv is calculated from vestibular motion input is referred to as SVC part. The SVC part is adopted from one of the version of 6DOF-SVC model (Inoue et al., 2023). The MISC is calculated from the subjective vertical in the MIS output part. In the present study, four different types of MISC output part (shown in Fig.2) are introduced and compared.

The outputs of the SVC part is the subjective vertical conflict $\Delta v$, which is defined as the sensed vertical ($v_s$) and the expected one ($\hat{v}_s$) calculated in the internal model. The inputs of the SVC part are Gravitoinertial acceleration (GIA) of the head in the three-dimensional space $f \in R^3$, which is defined as $f = a + g$, where $a$ and $g$ denote the inertial and gravitational accelerations, respectively, while $\omega \in R^3$ is the head angular velocity.

OTO and SCC denote the dynamics of the otolith and semicircular canals, respectively, while $\overline{OTO}$ and $\overline{SCC}$ represent the internal models in the CNS.

In the present study, OTO and $\overline{OTO}$ were set to the unit matrix, and the transfer functions identified in previous studies (Merfeld et al., 1993) were used for the SCC and $\overline{SCC}$ as follows:

$$\omega_s = \frac{\tau_d s}{\tau_d s + 1}\omega, \quad \hat{\omega}_s = \frac{\tau_d s}{\tau_d s + 1}\hat{\omega} \tag{1}$$

where the vectors with subscripts, such as $\omega_s$, denote afferent signals obtained from the sensory organs or expected signals obtained from the internal model.

According to Einstein's equivalence principle, otoliths receive GIA $f$, but not $a$, directly. In order to obtain $a$ and $g$, some calculation is required. For the GIA resolution, illustrated by the LP block in Fig.1, the generalized Mayne equation (Bos & Bles, 2002) is used:

$$\frac{dv_s}{dt} = \frac{1}{\tau}(f_s - v_s) - \omega_s \times v_s, \tag{2}$$

where $v_s$ denotes the gravity acceleration vector obtained from the sensory organs. Similar calculations were performed in the internal model.

Vectors representing discrepancies between the sensed and expected ones by the internal model $\Delta a = a_s - \hat{a}_s, \Delta \omega = \omega_s - \hat{\omega}_s, \Delta v = v_s - \hat{v}_s$ are fed back to the respective internal model of the sensory organs to obtain better self-motion perception.

Given the above, input to the internal model blocks are given as

$$\hat{f} = \tilde{a} + K_{ac}\Delta a + K_{vc}\int \Delta v(\tau)d\tau, \quad \hat{\omega} = \tilde{\omega} + K_{wc}\Delta\omega. \tag{3}$$

The vectors $\tilde{a}$ and $\tilde{\omega}$ denote signals that are the summation of multiple sources of information that contribute to self-motion perception, such as effect copy, information from other senses (i.e., somatosensory), and the effect of motion prediction (Wada, 2021). In the present study, signals are generated by multiplying the gains by the physical motions (Kamiji et al., 2007)(Wada et al., 2018) for the sake of simplicity.

B. MISC output part

The MISC output part calculates the MISC and symptom progression from the subjective vertical conflict $\Delta v$. Regarding the dynamics connecting sensory conflict to a generic symptom level, Oman proposed a structure that combines two parallel, interacting, critically damped second-order transfer functions with fast and slow dynamics(Oman, 1990). In addition, power scaling is used in the output of the dynamics to calculate generic symptoms in(Oman, 1990). A modified version of the model was developed to describe individual differences by introducing parameters to be tuned in fast and slow pathways(Irmak et al., 2021).

Based on the above, the present study introduces four types of MISC output parts as follows (Fig. 2).

**[(a) MSIbase]** is composed of a combination of the Hill function and a second-order transfer function, used in the conventional SVC models that outputs MSI as (Bos & Bles, 1998; Inoue et al., 2023; Kamiji et al., 2007).

**[(b) OmanAP]** has the same structure as in prior research(Irmak et al., 2021; Oman, 1990). The power-scaling part was placed after the fast and slow pathways.

**[(c) OmanBP]** has the same structure as that used in previous research. The power-scaling part is placed before the fast and slow pathways (Oman, 1990).

**[(d) OmanHILL]** is a model in which the input power scaling of OmanBP is replaced by the Hill function used in a prior study(Bos & Bles, 1998). We newly have introduced this structure as a natural expansion of the conventional SVC model by replacing the final path for MSI with fast and slow pathways.

The dynamics of the fast and slow pathways used in OmanAP, OmanBP, and OmanHILL in common are given as follows (also see Fig.2-(b)):

$$u_o = u_s + u_f, \quad u_s = \frac{1}{(\beta_2 s + 1)^2} u_i, \quad u_f = \frac{1}{(\beta_1 s + 1)^2} u_s u_i \qquad (4)$$

Input $u_i$ and output MISC for three models are defined as follows:

$$\text{OmanAP:} \quad u_i = \|\Delta v\|, \quad MISC = (u_o)^{M_{AP}} \qquad (5)$$

$$\text{OmanBP:} \quad u_i := (\|\Delta v\|)^{M_{BP}}, \quad MISC = u_o \qquad (6)$$

$$\text{OmanHILL:} \quad u_i = \frac{(\|\Delta v\|/b)^2}{1 + (\|\Delta v\|/b)^2}, \quad MISC = u_o \qquad (7)$$

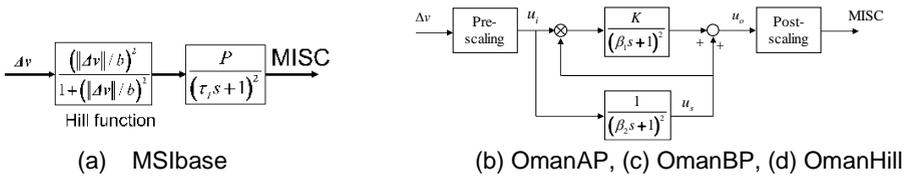

(a) MSIbase  (b) OmanAP, (c) OmanBP, (d) OmanHill

**Figure 2:** MISC output parts.

MODEL PARAMETER

We assumed that individual differences in motion sickness susceptibility is mainly determined by the sensitivity of the MISC output part. The parameters for the SVC were as follows: $K_a$=0.1, $K_w$=0.1, $K_{ac}$=0.5, $K_{wc}$=10.0, $K_{vc}$=5.0, $\tau$=2.0, and $\tau_d$=7.0, which were adopted from the In-1 model in(Inoue et al., 2023). The four different types of MISC output parts had three or four model parameters (Fig.2), all of which were positive and $\beta_1 < \beta_2$. These parameters were determined for each experimental condition and participant by minimizing the function (8):

$$J := \sum_{j=1}^{conN} \sum_{i=1}^{timeN} (MISC_{j,i}^{obs} - MISC_{j,i}^{mdl})^2 \qquad (8)$$

where $MISC_{j,i}^{obs}$ and $MISC_{j,i}^{mdl}$ are the MISC observed in the experiment and those calculated using the computational model, respectively. The scalars $conN (= 2)$ and $timeN (\leq 28)$ denote the numbers of experimental conditions used.

## EXPERIMENT

### APPARATUS

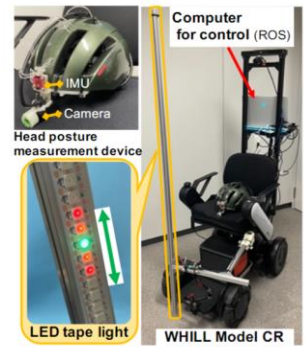

**Figure 3:** Apparatus.

An electric wheelchair WHILL model CR (WHILL Co., Ltd.) was automated by us and used in the experiment (Fig.3). An LED tape light attached to the wheelchair comprised multiple LEDs spaced 7 mm apart, each capable of changing color independently. Among the LEDs, only one was illuminated green at any given time. This specific LED could change over time so that the participants' head postures were guided to the GIA if they pursued the currently illuminated green LED using head movement without eye movement. The two LEDs positioned directly above and below the green LED are indicated were illuminated in red. An IMU was attached to the helmet to measure the acceleration and angular velocity acting on the heads of the participants. The WHILL moved in the fore-aft direction at a maximum speed of 1.67 m/s over a distance of 3m.

### DESIGN

Eleven participants were exposed to the fore-aft motion of an automated wheelchair while seated. The head-tilt condition, which was either move or static, was considered the primary experimental factor. In the movement condition, participants were instructed to pursue a green LED by tilting their heads with minimum eye movement. The position of the illuminated LED was determined such that the head was aligned with the GIA. In the static condition, they were instructed to gaze at the illuminated LED that was not moved so that the head was fixed in the Earth's vertical direction. Each participant performed two different head tilts on two different days, with the minimum interval of two days between the two experiments. The order of conditions was almost counterbalanced.

### PROCEDURE

This study was approved by the Ethics Review Committee of Nara Institute of Science and Technology, Japan (approval number: 2021-I-34). All participants provided written informed consent to participate after receiving an explanation of the experiment. Each participant was seated in an automated wheelchair in an upright posture and wore a helmet and seatbelt. The geometric relationship between the eye and LEDs was adjusted to accommodate individual differences among participants. Each participant was subsequently exposed to four consecutive sets of fore-aft linear wheelchair movements. The set was defined as a 5-minute movement of the wheelchair movement. A break of approximately 30-seconds was left between sets, during which the position of the wheelchair was initialized. During the motion exposure, the participants orally reported the MISC every minute. It should be noted that the motion challenge was stopped if the MISC reached 6, indicating the onset of nausea, or if the participants wanted to stop even when the four sets of the motion challenge were not completed. After the motion challenge, the participants were asked to remain seated in a stationary wheelchair to observe recovery from motion sickness every minute for 5 min.

PARTICIPANTS

Eleven participants (nine males, two females), with a mean age of 25.4 years (range 22–36 years), were recruited for the study. The percentiles of the motion sickness susceptibility questionnaire (MSSQ) score measured by the short version of the MSSQ (Golding, 2006)ranged from 0% to 85.2%, with a mean of 48.5% and a standard deviation of 30.4%.

## RESULT

OBSERVED MISC

Fig. 4 shows the time history of the observed MISC for each participant and each head tilt condition, together with the model-predicted MISC using the head movements measured during the experiment. Three participants who exhibited no symptoms (MISC=0) during the experiment were excluded from Fig.4.

MODEL-PREDICTED MISC

Time series of the MISC was predicted by inputting the head IMU data to the proposed models with optimized parameters (Fig.4).

Correlation analysis using the Pearson correlation coefficient between the observed and model-predicted MISC for all participants, excluding the three participants who exhibited no sickness throughout the experiment, showed very strong positive correlations in the OmanBP (0.92), OmanAP (0.90), and OmanHILL (0.93) models, whereas a strong correlation was observed in MSIbase (0.81). All correlations were statistically significant (p<0.000).

The mean absolute error of the observed and model-predicted MISC, defined in Eq.(9) for the four output conditions is shown in Fig. 5.

$$\text{mean absolute error} = \frac{1}{timeN} \sum_{i=1}^{timeN} | MISC_i^{Exp} - MISC_i^{mdl} | \qquad (9)$$

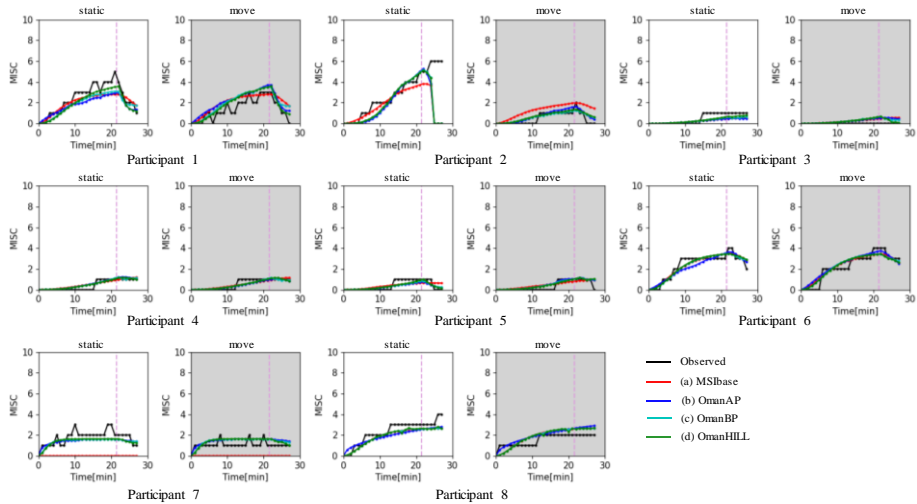

**Figure 4:** MISC observed for each participant under each condition. The model-predicted MISCs are also plotted. Three participants whose MISCs were zero for both head-tilt conditions were excluded.

A one-way repeated-measures analysis of variance (ANOVA) for the mean absolute error revealed the significance of the main effects of the output factor ($F(3,45) = 3.84$, $p = .016$). The post-hoc test using the paired t-test with Holm correction revealed that errors with MSIbase were significantly larger than those with OmanBP ($p<.0029$) and OmanHILL ($p<.00037$), and that errors with OmanAP were significantly larger than those with OmanHILL, while no significant difference was found between the other combinations.

## DISCUSSION AND CONCLUSION

The very strong positive correlations in the correlation analysis and the small mean absolute error for the three types of output parts with the Oman dynamics indicate that the proposed method combined with the Oman dynamics can accurately represent individual MISC changes, regardless of whether scaling is applied before or after the dynamics. Statistical analysis results indicated that OmanHILL and OmanBP were the most promising; however, it should be noted that comparing them under various scenarios is essential in the future.

A study (Irmak et al., 2021) previously proposed a model to describe individual differences in MISC by directly connecting the Oman dynamics used in the present study to the observed acceleration signals, rather than sensory conflict. Therefore, the contribution of the present study is the first to develop a model that calculates individual MISC based on the SVC theory and validates its effectiveness. The In1 model used in the SVC part of the present study was also partially validated for human perception in the direction of gravity, making it the first model capable of representing both motion perception and individual MISC. This characteristic is expected to play an important role in future additions such as the incorporation of visual inputs.

This study has several limitations which should be mentioned. Firstly, the motion scenarios tested in this study were limited. As such, the validation of more extensive motion scenarios, including rotation, is required. Additionally, the same data were used for model parameter identification and validation, suggesting that although the descriptive capability of the proposed method was confirmed, it is unclear whether the parameters obtained could be applied to other scenarios. This is also practically significant and requires further verification. Furthermore, this study focuses on vestibular motion sickness. Expansion to models that can address vestibular-visual interaction remains a future research direction. Note that the results in the present study, as well as such expansion, have been partially included in a Master's thesis, which is currently under embargo (Inoue, 2023).

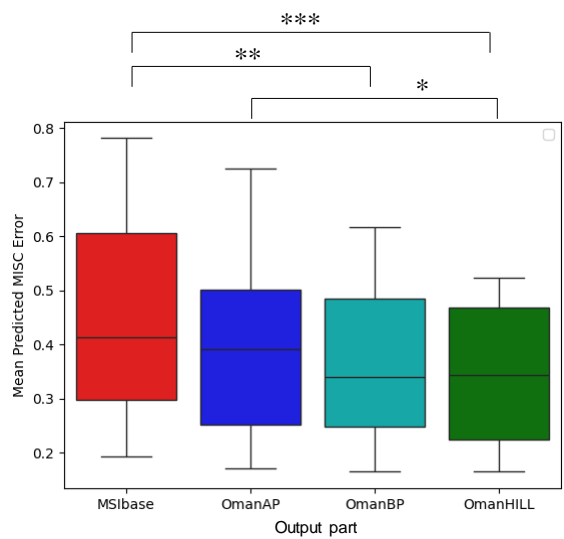

**Figure 5:** Mean absolute error of the observed and predicted MISC.
(*:p<.05, **: p<.01, ***:p<.001)

## ACKNOWLEDGMENT

This work was supported by JSPS KAKENHI (Grant Number 21K18308, 24H00298), Japan.